\documentclass[a4paper,fleqn]{cas-sc}

\usepackage[authoryear]{natbib}
\usepackage{graphicx} 
\usepackage{float}
\usepackage{algorithm}  
\usepackage{algpseudocode}
\usepackage{color}
\usepackage{setspace}
\usepackage[nomarkers,figuresonly]{endfloat}

\def\tsc#1{\csdef{#1}{\textsc{\lowercase{#1}}\xspace}}
\tsc{WGM}
\tsc{QE}
\tsc{EP}
\tsc{PMS}
\tsc{BEC}
\tsc{DE}

\begin{document}
\let\WriteBookmarks\relax
\def\floatpagepagefraction{1}
\def\textpagefraction{.001}
\shorttitle{Calculation of effective elastic properties using GPUs}
\shortauthors{Y. Alkhimenkov}

\title [mode = title]{Digital rock physics: calculation of effective elastic properties of heterogeneous materials using graphical processing units (GPUs)}

\author[1]{Yury Alkhimenkov}[type=editor,
                        auid=000,bioid=1,orcid=0000-0002-6529-9256]
\credit{Conceptualization, Methodology, Software, Validation, Writing - Original Draft, Writing - Review $\&$ Editing, Visualization, Project administration, Funding acquisition  }

%

\address[1]{Department of Civil and Environmental Engineering, Massachusetts Institute of Technology\\
77 Massachusetts Ave, Cambridge, 02139, MA, USA}

\begin{abstract}
An application based on graphical processing units (GPUs) applied to 3-D digital images is described for computing the linear anisotropic elastic properties of heterogeneous materials. The application can also retrieve the property contribution tensors of individual inclusions of any shape. The code can be executed on professional GPUs as well as on a basic laptop or personal computer Nvidia GPUs. The application is extremely fast: a calculation of the effective elastic properties of volumes consisting of about 7 million voxel elements ($191^3$) takes less than 4 seconds of computational time using a single A100 GPU; 3 minutes for 100 million voxel elements ($479^3$) using a single A100 GPU; 14 minutes for 350 million voxel elements ($703^3$) using a single A100 GPU. Several comparisons against analytical solutions are provided. In addition, an evaluation of the anisotropic effective elastic properties of a 3-D digital image of a cracked Carrara marble sample is presented. The software can be downloaded from a permanent repository Zenodo, the link with a doi is given in the manuscript.

\end{abstract}

\maketitle 

\printcredits

\doublespacing


\section{Introduction}
\label{intro}


The accurate calculation of effective elastic properties is pivotal across various fields, including mechanics, engineering, manufacturing, and geophysics. Traditionally, the focus has been on ellipsoidal inclusions and cracks, where analytical methods such as those developed by \cite{eshelby1957determination} have dominated. These methods have been refined over the decades to address heterogeneous composites \citep{mori1973average, benveniste1987new, shermergor1977Theory, willis1983overall, buryachenko2001multiparticle, nemat2013micromechanics}, and comprehensive reviews of these methodologies can be found in numerous monographs, such as those by \cite{kachanov2018micromechanics}. However, these traditional approaches are predominantly based on the Eshelby solution, limiting their application to ellipsoidal inhomogeneities and necessitating alternative methods for non-ellipsoidal shapes to accurately obtain the effective elastic properties and their property contribution tensors.


To address these complexities, homogenization theory has provided a robust mathematical framework, aiding in the understanding of the microstructural behavior of composite materials \citep{sanchez1980non, bakhvalov2012homogenisation, papanicolau1978asymptotic}. A review of computational homogenization is provided by \cite{otero2018multiscale}. Some key theoretical aspects of continuum micromechanics are given by \citep{zaoui2002continuum}.  A review of computational homogenization methods based on a Fourier transform is given by \cite{schneider2021review}. Different numerical approaches have been developed for the calculation of effective elastic properties. They include the finite element method \citep{garboczi1995algorithm, garboczi1998finite, roberts2002computation, arns2002computation}, dynamic wave propagation method \citep{saenger2007finite, saenger2008numerical}, the numerical method based on the Lippmann-Schwinger
equation \citep{moulinec1998numerical}, the finite element method included in the COMSOL software \citep{multiphysics1998introduction}, the spectral element method included in the FIDESYS software \citep{levin2019numerical}. \cite{drach2016comparison} analyzed the accuracy of full field and single pore analytical approaches using the numerical methods. \cite{markov2020unified} used a numerical method based on a class of Gaussian approximating functions to calculate the property contribution tensors. While all these methods have advantages and disadvantages, they are executed using a central processing unit (CPU). Some of these methods were applied to calculate effective elastic properties of 3-D images of rock samples \citep{andra2013digital, saxena2016estimating}. The strain energy-based method to estimate effective elastic properties was considered in \cite{zhang2007using}.


The last decade has ushered in a transformative shift in high-performance computing, driven by advancements in graphical processing units (GPUs). Nowadays numerical performance is governed by a memory access speed instead of the speed of floating-point calculations. Graphical processing units (GPUs) are many-core processors that can have a memory bandwidth of about one or two orders of magnitude higher than the memory bandwidth of a central processing unit (CPU). This shift has spurred the development of GPU-specific applications, which have shown substantial improvements in efficiency and capability. A high efficiency GPU-based Stokes solver that is applied to the incompressible flow in porous media is presented by \cite{evstigneev2023stationary}. GPU-based algorithms have been effectively utilized for modeling reactive solitary waves in three dimensions \citep{omlin2017pore}, viscoelastic deformation coupled with porous fluid flow \citep{omlin2018simulation, rass2019resolving}, thermomechanical coupling \citep{duretz2019resolving}, and thermomechanical ice deformation in both two and three dimensions \citep{rass2020modelling}. A simulation of wave propagation in poroelastic media, resolving over 1.5 billion grid cells in just a few seconds was presented by \cite{alkhimenkov2021resolving, alkhimenkov2021stability}. Additionally, an efficient numerical implementation of pseudo-transient iterative solvers on GPUs for quasi-static problems has been achieved using the Julia programming language \citep{rass2022assessing}. Recently, compaction‐driven fluid flow and plastic yielding within porous media was investigated through numerical modeling \citep{alkhimenkov2024shear}. 
A numerical approach based on graphical processing units (GPU) to resolve the strain localization in two and three dimensions of a (visco)-hypoelastic-perfectly plastic medium was developed by \cite{https://doi.org/10.1029/2023JB028566}. 

In this contribution, an efficient application based on graphical processing units (GPUs) is presented for the calculation of effective elastic properties of heterogeneous materials. The application can also be adopted to calculate the property contribution tensors of inclusions of any shape. A solution of the quasi-static equations is performed using the Accelerated Pseudo-transient method which is local and can be naturally parallelized \citep{rass2022assessing, alkhimenkov2024PT}. The present method is designed for GPUs, leveraging its multiprocessor power, and is approximately two orders of magnitude faster compared to the methods designed for CPUs. A special attenuation is given to the performance: the application is extremely fast. Effective elastic properties of models involving more than 100 million voxel elements can be calculated in less than 4 minutes using a professional GPU. The application can be executed on personal computers or laptops which feature an Nvidia GPU. Several comparisons against analytical solutions are provided. In addition, an evaluation of the anisotropic effective elastic properties of a 3-D digital image of a cracked Carrara marble sample is presented. The routines are available from a permanent DOI repository (Zenodo) \url{https://doi.org/10.5281/zenodo.13998253} \citep{alkhimenkov_2024_13998253} [Software].


\section{Theory}

\subsection{Closed System of Equations}
Consider a volume $V$ in a three-dimensional Euclidean space $E^3$ bounded by a regular surface $\partial V$ and a point in $V$, which is identified by its position vector $\pmb{x}$ with components $x_p$ ($p \, = \, \overline{1,..,3}$). Let us consider a fixed rectangular Cartesian coordinate system. The conservation of linear momentum or the equilibrium equation is \citep{landau1959courseElast, nemat2013micromechanics} 
\begin{equation}\label{eq:1}
\pmb{\nabla} \cdot  {\pmb{ \sigma}} (\pmb{x}) = 0 ,
\end{equation}
where ${\pmb{ \sigma}}$ is the stress tensor, $\pmb{\nabla}$ is the del operator, $\cdot$ is the dot product and $ \pmb{\nabla}  \cdot$ denotes the divergence operator. In a rate formulation, the constitutive equation can be written as
\begin{equation}\label{eq:2} 
\dot{\pmb{ \sigma}} (\pmb{x})= \mathbf{C}(\pmb{x}) : \dot{\pmb{\varepsilon}} (\pmb{x} )  
\end{equation} 
where $:$ is the double dot product, $\dot{\pmb{ \sigma}}$ is the stress-rate tensor, $\dot{\pmb{\varepsilon}}$ is the strain-rate tensor and $\textbf{v}$ is the velocity field. The strain-rate-velocity relation is \citep{nemat2013micromechanics, li2008introduction}
\begin{equation}\label{eq:3}
\dot{\pmb{\varepsilon}} (\pmb{x})  = \frac{1}{2} \left( \pmb{\nabla } \otimes\textbf{v}  + (\pmb{\nabla}  \otimes\textbf{v} )^\text{T} \right),
\end{equation}     
where $\otimes$ is the tensor product, the superscript ``$^\text{T}$" denotes transpose. In a component form, equations \eqref{eq:1}-\eqref{eq:3} can be rewritten as
\begin{equation}\label{eq:1a}
\nabla_j \sigma_{ij}(x_p) = 0,
\end{equation}
\begin{equation}\label{eq:2a}
\dot{\sigma}_{ij} (x_p) = C_{ijkl}(x_p) \dot{\varepsilon} _{kl}(x_p),
\end{equation} 
\begin{equation}\label{eq:3a}
\dot{\varepsilon} _{ij}(x_p) = \frac{1}{2} \left( \nabla_i v_j  + \nabla_j v_i  \right) ,
\end{equation}  
where $i,j,k,l=\overline{1..3}$ and Einstein summation convention is applied, i.e. summation is applied over repeated indexes. In this study, the infinitesimal strain theory is adopted. No distinction is made between
the Lagrangian and the Eulerian descriptions. All quantities are expressed in
terms of $\pmb{x}$ (with components $x_p$, $p \, = \, \overline{1,..,3}$) which is interpreted as the initial particle position \citep{nemat2013micromechanics}.

\subsection{Effective elastic properties}

Let us derive the macro-fields namely the volume-averaged stress-rate and strain-rate tensors over the volume $V$. The volume-averaged stress-rate tensor can be written as \citep{nemat2013micromechanics}
      \begin{equation}\label{BC:2}
    \langle  { {\sigma } }_{ij}  \rangle = \frac{1}{V} \int _V  { {\sigma} }_{ij} ( {x}_p  ) {dV}
    \end{equation}
and the strain-rate tensor is \citep{nemat2013micromechanics}   
  \begin{equation}\label{BC:1}
    \langle  { {\varepsilon}}_{kl}  \rangle = \frac{1}{V} \int _V  { {\varepsilon}}_{kl} ({x}_p ) {dV} ,
    \end{equation}\newline
    where $\langle \cdot \rangle$ represents the volume averaging over the numerical domain $V$. The effective stiffness tensor ${\text{C}}^*_{ijkl}$ is now can be defined as:
      \begin{equation}\label{BC:3}
    \langle  { {\sigma }}_{ij}   \rangle =  {\text{C}}^*_{ijkl}   \langle  { {\varepsilon}}_{kl}   \rangle.
    \end{equation}\newline
If volume $V$ is a Representative volume element (RVE) of a particular material then ${\text{C}}^*_{ijkl}$ is effective stiffness tensor of this material. For simplicity, Voigt notation is adopted in this study as to represent the components of tensors, i.e., $C_{ijkl} = C_{mn}$, $i,j, k, l \, = \, \overline{1,..,3}$ and $m,n\, = \, \overline{1,..,6}$.

\section{Numerical solution strategy}

\subsection{Solution of the system of equations}

The solution of the inertialess equations \eqref{eq:1a}-\eqref{eq:3a} is performed using the matrix-free accelerated pseudo-transient (APT) method \citep{frankel1950convergence, rass2022assessing, alkhimenkov2024PT}. The main idea of this method is that instead of solving inertialess equations, dynamic (hyperbolic) equations are solved with an appropriate attenuation of the dynamic fields. Once the dynamic fields attenuate to a certain precision (for example, to the $10^{-12}$), the solution of the inertialess equations is achieved. In other words, the quasi-static problem is an attractor of the hyperbolic one with damping. The APT method can be used to solve numerical domains evolving more than a billion voxel elements. Another advantage of the APT method is that all operations a local and, thus, the method can be naturally parallelized using, for example, GPUs.

In the numerical solver, the stress tensor, $\sigma_{ij}$, is partitioned into the pressure, $p$, and the stress deviator, ${\tau} _{ij}$:
\begin{equation}\label{A2ssP}
\sigma_{ij} =  - {p} \delta _{ij} +  {\tau} _{ij},
\end{equation}
where $\delta _{ij}$ is the Kronecker delta. The solution of the quasi-static elasticity equations \eqref{eq:1a}-\eqref{eq:3a} can be achieved in two steps. First, inertial terms are added into the equation for the stress; second, a Maxwell rheology (viscous damper) is added to equations for the stress deviator. The equilibrium equation \eqref{eq:1a} can be re-written with the pseudo-time $\widetilde{t}$,
\begin{equation}\label{11111}
\nabla _j \, \left( - {p} \delta _{ij} +  {\tau} _{ij} \right) = \widetilde{\rho} \dfrac{\partial v_i}{\partial \widetilde{t}}.
\end{equation}

The pseudo-transient version of the equation for the pressure ${p}$ becomes:
\begin{equation}\label{A2ssP}
\frac{1}{\widetilde{K} } \frac{\partial  {p}}{\partial \widetilde{t}}  +  \frac{1}{K} \frac{  {p}  }{\Delta t}= -  \nabla_k v_k . 
\end{equation}
For the stress deviator the corresponding equation is
\begin{equation}\label{A23ssP2}
\frac{1}{2 \widetilde{G} } \frac{\partial{} \tau_{ij}}{\partial{} \widetilde{t}}  +    \frac{1}{2 G} \frac{ \tau_{ij}  }{\Delta t}+ \frac{\tau_{ij}}{2 \mu_s}  =   \frac{1}{2} \left( \nabla_i v_j + \nabla_j v_i - \frac{2}{3}\nabla_k v_k \right),
\end{equation}
where $\mu_s$ is the viscosity of the solid, which is set to a very high value (e.g., $\mu_s=10^{20}$) or can be neglected. In equations \eqref{11111}-\eqref{A23ssP2}, the quantities $\widetilde{K}  \partial{} \widetilde{t}$, $\widetilde{G}_1 \partial{} \widetilde{t}$ and $ \widetilde{\rho}/\partial{} \widetilde{t}$ are the numerical parameters. The optimal values for these parameters can be determined numerically or analytically. In the present study, the following expressions are used as outlined in \cite{alkhimenkov2024PT}. The Courant–Friedrichs–Lewy (CFL) condition for the system of equation \eqref{11111}-\eqref{A23ssP2} suggest that \citep{alkhimenkov2021stability}
\begin{equation}\label{PT1_1t2}
\partial \widetilde{t} \leq \dfrac{\Delta x}{\widetilde{V}_p} \qquad\textnormal{or}\qquad  \partial \widetilde{t} = \dfrac{\widetilde{C} \Delta x}{\widetilde{V}_p},
\end{equation}
where $\widetilde{C} \leq 1$ and the ``numerical" primary or P-wave velocity can be calculated as:
\begin{equation}\label{PT1_1t}
\widetilde{V}_p = \sqrt{\dfrac{\widetilde{H}  }{\widetilde{\rho}}},
\end{equation}
Note that $\widetilde{H} = \widetilde{K} + \frac{4}{3}\widetilde{G} $ and $\widetilde{K}=K $, $\widetilde{G}=G$. The dimensionless parameter, the Strouhal number, ${\mathrm{St}}$, is expressed as 
\begin{equation}\label{PT1_1t311}
 {\mathrm{St}}=  \dfrac{  L_x}{\widetilde{V}_p  \, \Delta t} ,
\end{equation}
where $L_x$ is the physical size of the model in x-dimension, $\Delta t$ is the physical time step. The first numerical combination can be written as
\begin{equation}\label{PT1_1t31222}
\dfrac{\partial \widetilde{t} }{ \widetilde{\rho} } =  \dfrac{\widetilde{C} \Delta x \,L_x}{\mathrm{St} \,\widetilde{H} \, \Delta t}.
\end{equation}

The second numerical combination can be written as
\begin{equation}\label{PT1_1t31333}
  \widetilde{H}  \partial\widetilde{t} = (\widetilde{V}_p   \, \partial \widetilde{t})^2 \left( \dfrac{ \partial    \widetilde{t}}{\widetilde{\rho}} \right)^{-1}, 
\end{equation}
Note that $\widetilde{V}_p   \, \partial \widetilde{t}$ and $\partial \widetilde{t} / \widetilde{\rho}$ are already defined above, therefore, it is straightforward to calculate $\widetilde{H}  \partial\widetilde{t}$. The stress tenor in decomposed into pressure and deviatoric stress tensor, therefore, the following expressions are also provided
\begin{equation}\label{PT1_1t31333A}
  \widetilde{G}  \partial\widetilde{t} = (\widetilde{V}_p   \, \partial \widetilde{t})^2 \left( \dfrac{ \partial    \widetilde{t}}{\widetilde{\rho}} \right)^{-1} \left( K_G + \frac{4}{3} \right)^{-1}  , 
\end{equation}
where $K_G = K/G$, and
\begin{equation}\label{PT1_1t31333AB}
\widetilde{K}  \partial\widetilde{t} = K_G \, \widetilde{G}  \partial\widetilde{t}.
\end{equation}
The dispersion and convergence analysis for the employed equations suggests the following optimal values for the numerical parameters \citep{alkhimenkov2024PT}:
\begin{equation}\label{PT24q}
\mathrm{St}_{opt} = 2\, \pi,  \qquad {K_G}^{opt} = \dfrac{K_1}{G_1 } , 
\end{equation}
where $K_1$ and $G_1$ are the bulk and shear moduli of the background material, respectively.
Numerical tests show that the provided values for $\mathrm{St}_{opt}$ and ${K_G}^{opt}$ remain valid in 1D, 2D, and 3D. However, in 2D, $\mathrm{St}_{opt}$ is multiplied by $\sqrt{2}$, and in 3D, $\mathrm{St}_{opt}$ is multiplied by $\sqrt{3}$. A typical number of iterations over the pseudo-time depends on the problem size (in grid cells), the convergence rate and the desired precision. Form our experiments, a typical 3D heterogeneous model requires from $5\times n_x$ to $10\times n_x$ ($n_x$ is the number of grid cells in x-dimension) iterations over the pseudo-time to achieve the quasi-static solution.


\subsection{Hardware}

The solver is implemented using the CUDA C language. Numerical results were calculated using two GPU systems: (i) Laptop Nvidia RTX 4090 16GB GPU and (ii) Nvidia A100 80GB professional GPU. The A100 GPU is approximately three times faster than the RTX 4090. The solver is quite simple, it consists of a single ``\texttt{for}" loop over a pseudo-time $\widetilde{t}$. The solution is reached once the dynamic fields attenuate to a desired precision. For simplicity, a dimensionless form of the equations is not introduced because the problem is already simple. From the theoretical point of view, the solver exhibits a first-order accuracy due to the application of the pseudo-transient method.

\subsection{Discretization}

\begin{figure}
\centering
\includegraphics[width=0.7\textwidth]{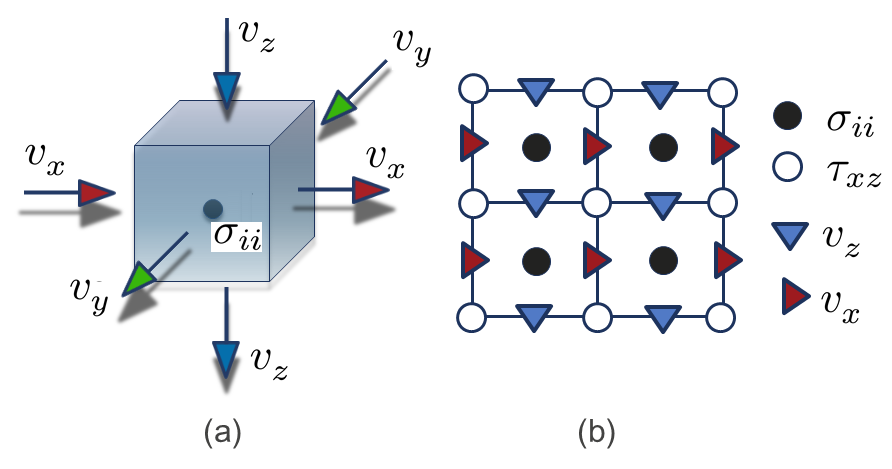}
\caption{A sketch representing (a) the finite volume, where the
velocities preserve mass balance and (b) the spatial positions of stress and velocity fields in the X–Z plane [modified after \cite{alkhimenkov2021resolving}].}
\label{Figure52}%
\end{figure}

The numerical domain $V$ is discretized using a staggered in space and time grid \citep{virieux1986p}. This approach provides us with a variant of a conservative finite volume approach \citep{dormy1995numerical}. In other words, the numerical domain is partitioned into voxel elements. Each element can represent a background matrix material or an inclusion material. The total number of voxel elements is restricted only by the available GPU memory. Figure~\ref{Figure52} shows a schematic representation of spatial positions of stress and velocity fields. The employed discrete scheme exhibits second order accuracy in space and time. A detailed representation of discrete equations can be found in \cite{alkhimenkov2021resolving}.

\subsection{Boundary conditions and the direct tests }

To solve a particular partial differential equation (PDE) appropriate boundary conditions are needed. Such boundary conditions can be specified as Dirichlet, Neumann, or mixed boundary conditions, etc. In this applied study, PDEs are formulated in terms of stress- and strain-rate quantities, therefore, velocity boundary conditions $\Delta v $ are applied to the external walls of the numerical domain. For example, to calculate the ${\text{C}}^*_{33}$ component of the stiffness tensor, $\Delta v=10^{4}  $ is applied to a single boundary of the numerical domain in $z$ direction in a perpendicular direction. An opposite boundary is subjected to $\Delta v=0  $ boundary condition. Boundaries parallel to $x$ and $y$ directions are subjected to the reflecting boundary conditions (Dirichlet conditions). Once the solution is achieved, the volume-averaged stress and strain tensors are calculated according to equations \eqref{BC:2}-\eqref{BC:1} and the resulting $C_{33}(\omega_i)$ component of the effective stiffness tensor medium is calculated as 
\begin{equation}\label{eq:4}
C_{33}  = \frac{\langle \dot{\sigma}_{zz} \rangle}{\langle\dot{\epsilon}_{zz} \rangle }.
\end{equation}

In a similar way, the components $C_{11}$ and $C_{22}$ are obtained. Similar direct tests are used to calculate other components of the stiffness tensor. A full set of applied boundary conditions is given in Appendix~\ref{sec:sample:appendix}. In this study, it is assumed that the material belongs to, at least, an orthorhombic symmetry class (9 independent components). However, the presented method can be further extended to capture the most general anisotropic domains (i.e., the triclinic symmetry class).


\section{Numerical simulations}

To show the solver capabilities, simulations for three different models are presented: layered media,  an oblate spheroid embedded into an isotropic matrix and a 3-D digital image of a rock sample. 
Prepossessing, postprocessing and visualization are done in Matlab. To run the application, one needs to run the corresponding Matlab script which compiles and runs the CUDA C application on a GPU.


\subsection{Model 1: Validation for Layered Media}

\begin{figure}
\centering
\includegraphics[width=0.99\textwidth]{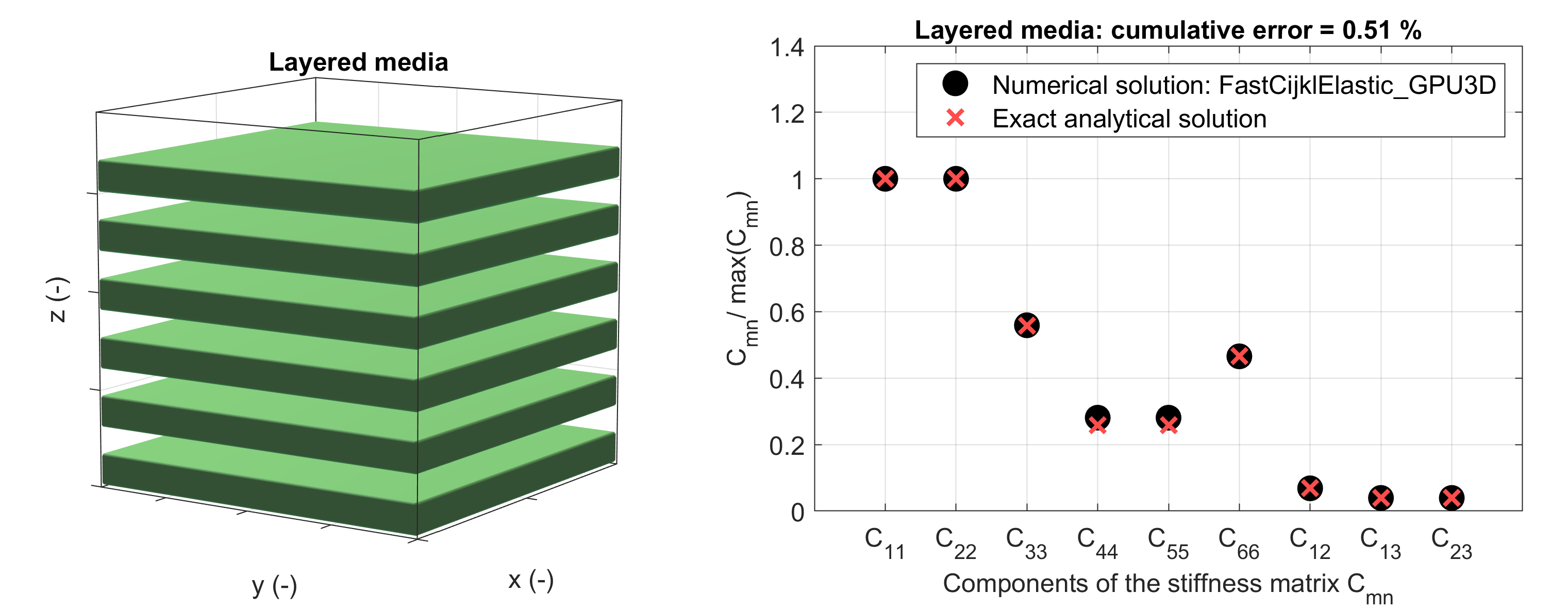}
\caption{Panel (a) shows the 3-D geometry of a layered media. Panel (b) shows the effective elastic properties calculated using the numerical solver (black circles) and the exact analytical solution (red crosses). Cumulative error between the numerical and analytical results is less then $0.1\%$.}
\label{L_v0}%
\end{figure}

First, let us consider a material consisting of several parallel isotropic layers (Figure~\ref{L_v0}a). For this geometry an exact analytical solution exists which is given by \cite{backus1962long}. The material parameters used are shown in Table~\ref{tbl1L}. The volume fraction of the stiff layers is $0.4712$.

For the calculation of the effective elastic properties, the numerical domain was discretized by $N=191\times191\times191=6,967,871$ voxels. The thickness of each layer is 15 voxels. The computation time for one component was $3-4$ seconds using a A100 GPU. Figure~\ref{Sph_v0}b shows the normalized components of the stiffness tensor calculated using the numerical solver (black circles) and the exact analytical solution (red crosses). The cumulative error is less than $0.6\%$ which confirms the accuracy of the GPU-based solver. Note that in this example such a low resolution is used on purpose --- to show the solver capabilities even for low-resolution 3-D images. The accuracy can be significantly improved if a high resolution 3-D image is used. The comparison between analytical and numerical results are shown in Table~\ref{tbl1Esh1}

{
\begin{table}[ ] 
\caption{Material properties of the layers}
\centering
\vspace{+1.0 mm}
 \begin{tabular}{| l | l | l |  }
  \hline			
  Material property & Layer 1  & Layer 2    \\
	\hline
	\hline
    Bulk modulus $K$       & 36\,GPa  & 7.2\,GPa   \\
  Shear modulus $G $   & 44\,GPa   & 8.8\,GPa       \\
  \hline  
 \end{tabular}
 \label{tbl1L}
\end{table}
}

{
\begin{table}[ ] 
\caption{Analytical and numerical results for layered media}
\centering
\vspace{+1.0 mm}
 \begin{tabular}{| l | l | l |  }
  \hline			
  $C_{kl}$ & Analytical solution  & Numerical solution    \\
	\hline
	\hline
    $C_{11}$       & 54.499\,GPa  & 54.499\,GPa   \\
  $C_{22} $   & 54.499\,GPa   & 54.499\,GPa       \\
  $C_{33} $   & 30.389\,GPa   & 30.419\,GPa       \\
  $C_{44} $   & 14.124\,GPa   & 15.307\,GPa       \\
  $C_{55} $   & 14.124\,GPa   & 15.307\,GPa       \\
  $C_{66} $   & 25.386\,GPa   & 25.376\,GPa       \\
  $C_{12} $   &  3.726\,GPa   & 3.720\,GPa       \\
  $C_{13} $   &   2.140\,GPa   & 2.148\,GPa       \\
  $C_{23} $   &  2.140\,GPa   & 2.148\,GPa       \\
  \hline  
 \end{tabular}
 \label{tbl1Esh1}
\end{table}
}
Note, that the discrepancy in $C_{44} $ and $C_{55} $ between analytical and numerical results is directly related to the boundary conditions imposed in the numerical calculations (there is a difference depending on how the boundary conditions are imposed: parallel or perpendicular to the layers).

\subsection{Model 2: validation against an Eshelby solution}

\begin{figure}
\centering
\includegraphics[width=0.99\textwidth]{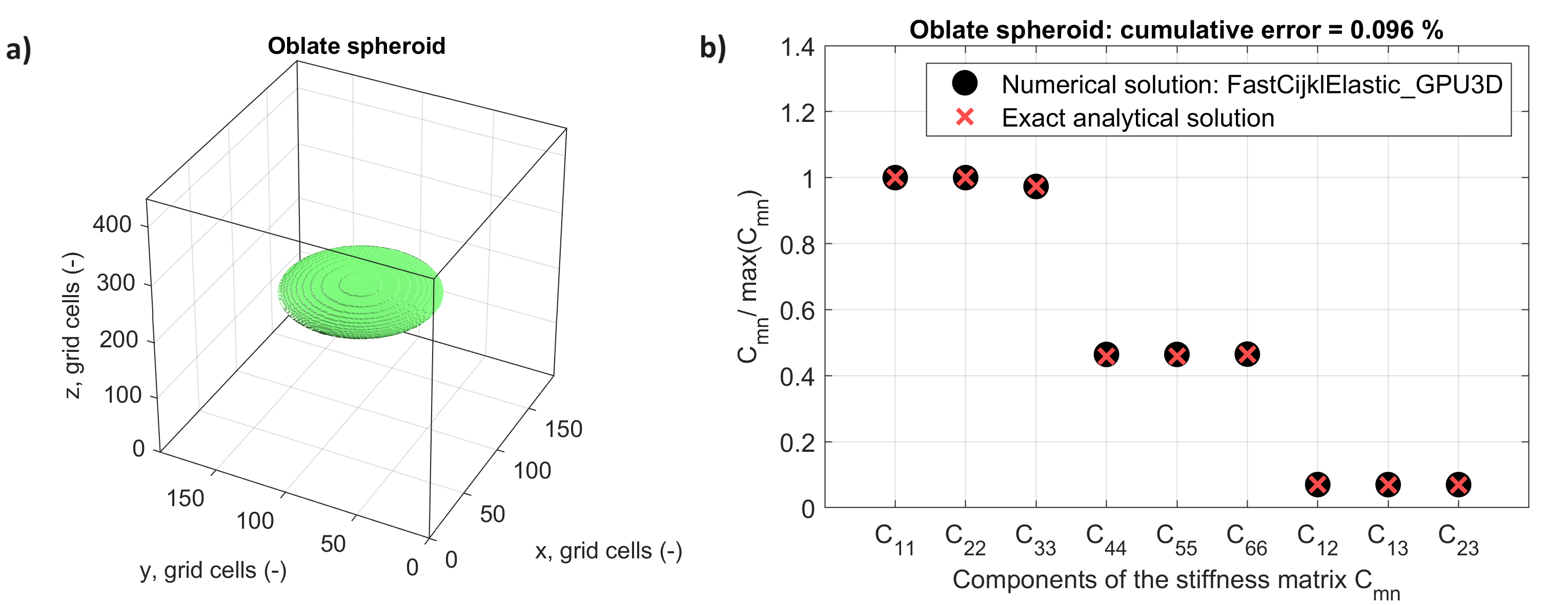}
\caption{Panel (a) shows the 3-D geometry of an oblate spheroid embedded into the isotropic material. Panel (b) shows the effective elastic properties calculated using the numerical solver (black circles) and the exact analytical solution (red crosses). Cumulative error between the numerical and analytical results is less then $0.1\%$.}
\label{Sph_v0}%
\end{figure}

The effective compliance tensor $\mathbf{S}^{*} = \left( \mathbf{C}^{*} \right)^{-1}$ (with components $S_{mn}^{*}$) for the material represented by a spheroid embedded into an isotropic matrix (Figure~\ref{Sph_v0}a) can be written as \citep{kachanov2018micromechanics}
\begin{equation}\label{eqN2}
\mathbf{S}^{*} = \mathbf{S}^{\textrm{1}} +  \phi \mathbf{H}   \qquad\textnormal{or}\qquad  S_{mn}^{*} = S_{mn}^{\textrm{1}} +  \phi H_{mn}  ,
\end{equation}
where $\mathbf{S}^{\textrm{1}}$ is is the compliance tensor (with components $S_{mn}^{\textrm{1}}$ ) of the background grain material, $\mathbf{H}$ is the compliance contribution tensor (with components $H_{mn}$) of the inclusion represented by a spheroid and $\phi$ is the inclusion volume fraction. The expression \eqref{eqN2} can be treated as approximation of non-interacting inclusions while tensor $\mathbf{H}$ (or matrix $H_{mn}$) is exact. The explicit form of the tensor $\mathbf{H}$ is based on the Eshelby solution \citep{eshelby1957determination} and is presented in~\ref{sec:sample:appendix2}. The material parameters used are shown in Table~\ref{tbl1Esh}. The inclusion volume fraction is $\phi=0.036$, the inclusion aspect ratio is $0.1$.

For the calculation of the effective elastic properties, the numerical domain was discretized by $N=191\times191\times447=16,307,007$ voxels. Due to the symmetry of the employed geometry, the actual calculation was performed for a quarter of the spheroid located in the corner of the numerical domain. The computation time for one component was $10-12$ seconds using a A100 GPU. Figure~\ref{Sph_v0}b shows the normalized components of the stiffness tensor calculated using the numerical solver (black circles) and the analytical solution (red crosses) via expression \eqref{eqN2}. The comparison between analytical and numerical results are shown in Table~\ref{tbl1Esh12}. The cumulative error is less than $0.1\%$ which confirms the accuracy of the GPU-based solver. Numerical tests show that the numerical solver can be used to calculate the effective elastic properties of models with inclusions with aspect ratios of $0.001$. In principle, the only limitation is the GPU memory, therefore, on professional GPUs (especially with a message passing interface, MPI) the aspect ratio can be further reduced. 

{
\begin{table}[ ] 
\caption{Material properties of the model with inclusion}
\centering
\vspace{+1.0 mm}
 \begin{tabular}{| l | l | l |  }
  \hline			
  Material property & Background  & Inclusion    \\
	\hline
	\hline
    Bulk modulus $K$       & 36\,GPa  & $10^{-3}$\,GPa   \\
  Shear modulus $G $   & 44\,GPa   & $2 \times 10^{-3}$\,GPa       \\
  \hline  
 \end{tabular}
 \label{tbl1Esh}
\end{table}
}

{
\begin{table}[ ] 
\caption{Analytical and numerical results for a single inclusion model}
\centering
\vspace{+1.0 mm}
 \begin{tabular}{| l | l | l |  }
  \hline			
  $C_{kl}$ & Analytical solution  & Numerical solution    \\
	\hline
	\hline
    $C_{11}$       & 94.425\,GPa  & 94.438\,GPa   \\
  $C_{22} $   & 94.425\,GPa   & 94.438\,GPa       \\
  $C_{33} $   & 91.929\,GPa   & 91.906\,GPa       \\
  $C_{44} $   & 43.283\,GPa   & 43.836\,GPa       \\
  $C_{55} $   & 43.283\,GPa   & 43.836\,GPa       \\
  $C_{66} $   & 43.888\,GPa   & 43.921\,GPa       \\
  $C_{12} $   &  6.6482\,GPa   & 6.6145\,GPa       \\
  $C_{13} $   &  6.5653\,GPa   & 6.5951\,GPa       \\
  $C_{23} $   &   6.5653\,GPa   & 6.5951\,GPa       \\
  \hline  
 \end{tabular}
 \label{tbl1Esh12}
\end{table}
}

\subsubsection{Compliance contribution tensor $H_{ijkl}$}

\begin{figure}
\centering
\includegraphics[width=0.7\textwidth]{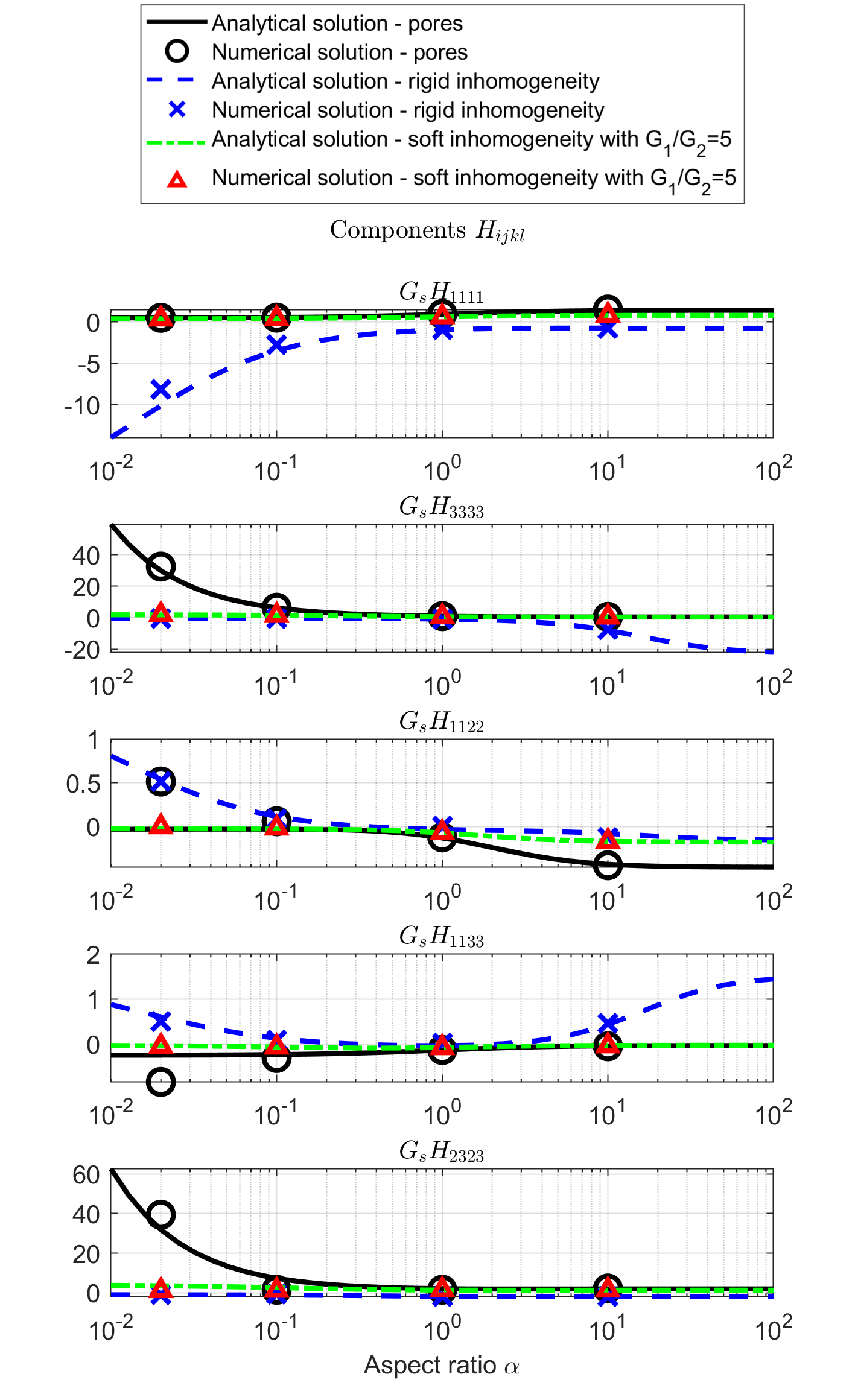}
\caption{Components $H_{ijkl}$ as a function of aspect ratio $\alpha$ and elastic contrast. $G_s$ denotes the shear modulus of the background medium, $G_1$ denotes the shear modulus of the host material and $G_2$ denotes the shear modulus of the inclusion. In addition, in the soft inhomogeneity experiment bulk moduli are $K_1/K_2 = 5$, where $K_1$ denotes the bulk modulus of the host material and $K_2$ denotes the bulk modulus of the inclusion.}
\label{22}%
\end{figure}

To further validate the solver, tensor $H_{ijkl}$ is calculated for various contrasts of the elastic moduli of the inclusion and the background medium and various aspect ratios of the inclusion (Figure~\ref{22}). It can be noted that the accuracy of the solver is sufficient for $\alpha \in [0.1; 10] $ for various contrasts of the elastic moduli of the inclusion and the background medium and reduces for off-diagonal components ($H_{1122}$, $H_{1133}$ ) for aspect ratio $\alpha= 0.02$. The accuracy can be greatly improved if the solver is executed with a higher resolution (lager than the currently employed resolution of 16 million voxel elements).


\subsubsection{Convergence rate}

\begin{figure}
\centering
\includegraphics[width=0.6\textwidth]{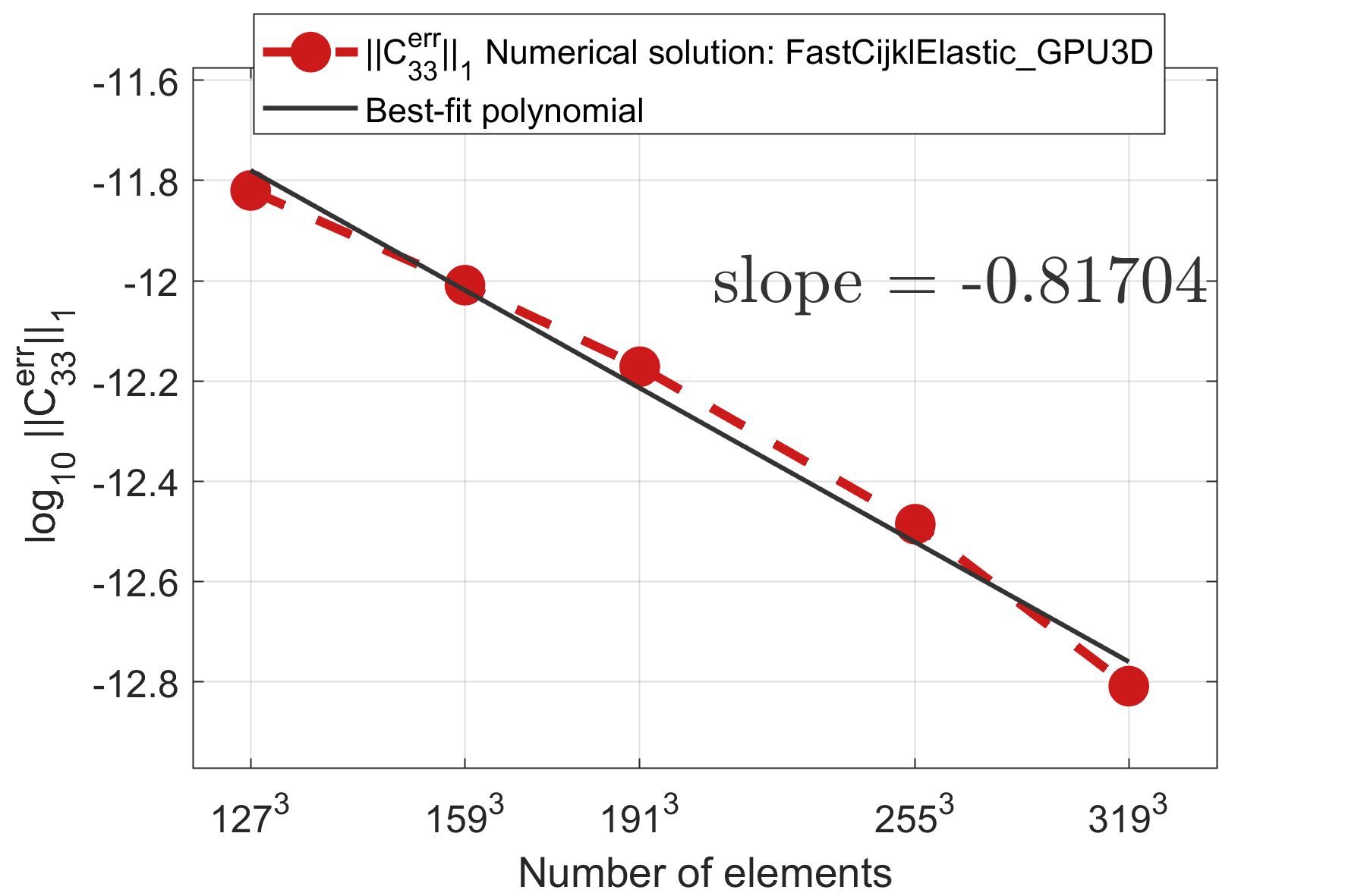}
\caption{Convergence of the numerical results as a function of voxel elements (in a logarithmic scale).}
\label{Conv}%
\end{figure}

\begin{figure}
\centering
\includegraphics[width=0.6\textwidth]{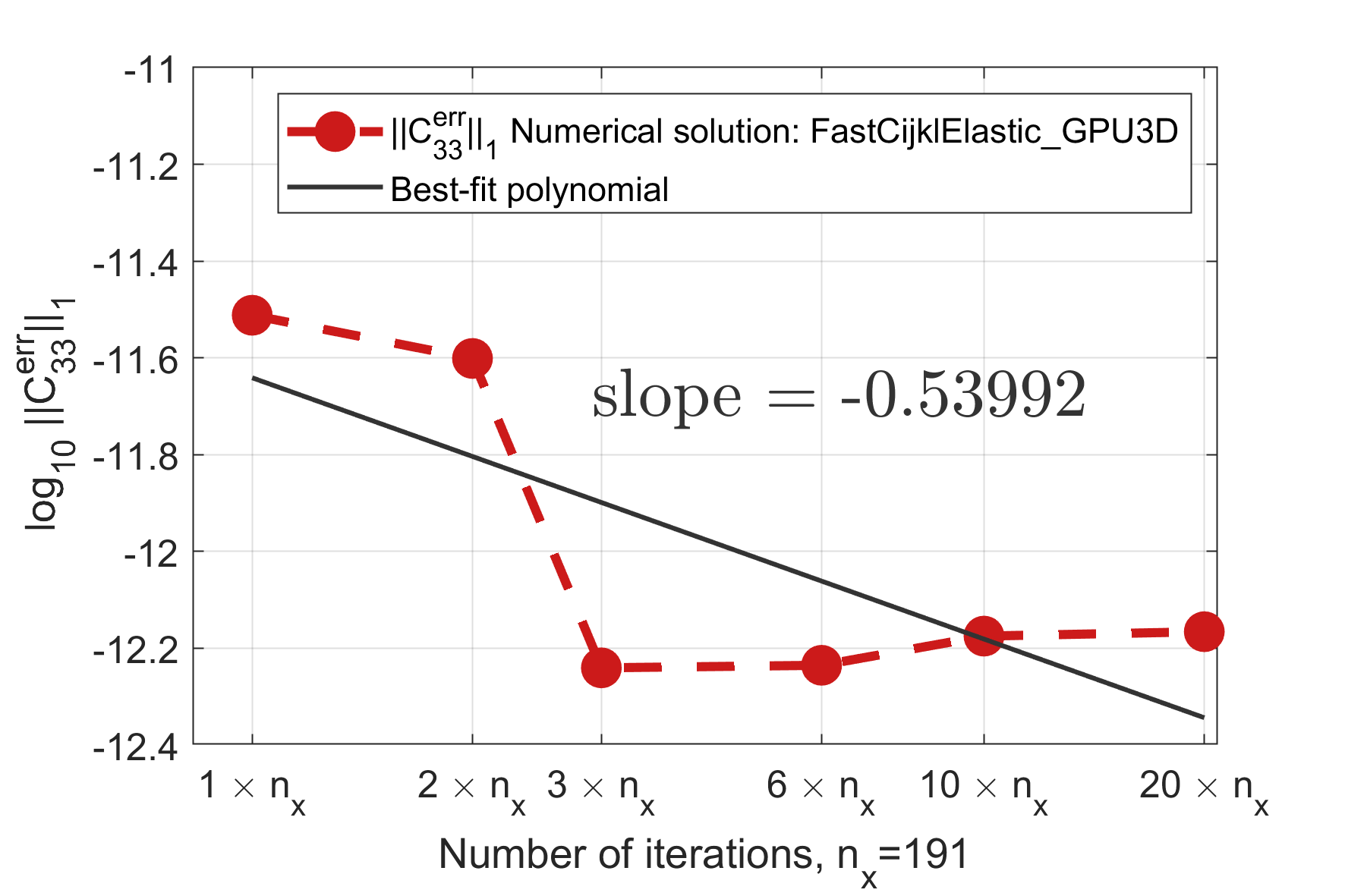}
\caption{Convergence of the numerical results as a function of voxel elements (in a logarithmic scale).}
\label{Conv_iter}%
\end{figure}

The convergence rate of the numerical solver as a function of numerical resolution is $0.81$, slightly lower than the expected value for first-order convergence (Figure~\ref{Conv}). This discrepancy arises because the analytical solution, assuming an infinite domain and no elastic iterations, is an approximation. Consequently, the numerical solution converges to a slightly different value, influenced by the domain's boundaries. Moreover, the convergence rate as a function of the number of iterations demonstrates that the numerical solution converges to a specific value, aligning with the analytical solution to several significant digits only (Figure~\ref{Conv_iter}).

\subsection{Model 3: elastic properties of a 3-D digital image of a rock sample}

\begin{figure}
\centering
\includegraphics[width=0.99\textwidth]{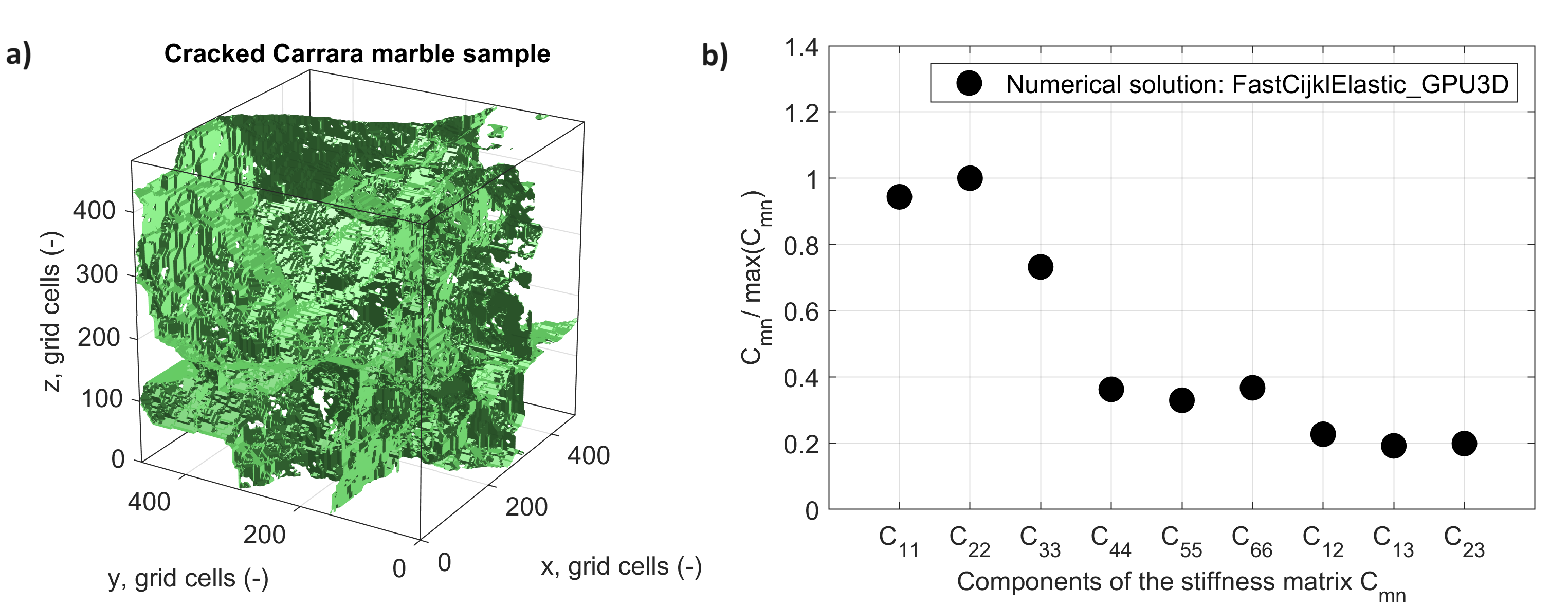}
\caption{Panel (a) shows the 3-D tomography Image of a cracked Carrara marble sample. Panel (b) shows the effective elastic properties calculated using the numerical solver (black circles). The resolution of the 3-D model is $N=703^3$.}
\label{R_v0}%
\end{figure}

\cite{ruf2020micro} published 3-D tomography Images of a cracked Carrara marble sample. The images were then segmented by \cite{lissa2021digital, simon_lissa_2021_4746139}. The resulting 3-D segmented image of a cracked Carrara marble is used as input into the solver --- to calculate the effective elastic properties. The 3-D image was first discretized using voxel elements. For that, an open access software is adopted \citep{AdamA}. Alternatively, other applications can be used, for example, \cite{MeshLib}, etc. Figure~\ref{R_v0}a shows the 3-D digital image of voxel elements of a cracked Carrara marble sample, green colors represent the dry pore space and the transparent domain represents the background isotropic material. The material parameters used are shown in Table~\ref{tbl1real}. The cracks volume fraction is $0.04$.

Several different resolutions were employed to study the convergence, which is another way to validate the numerical results. The computation time ranged from 3 seconds for the coarsest resolution $N=191^3$ to 14 minutes for the finest resolution $N=703^3$. Figure~\ref{R_v0}b shows the evaluated components of the effective stiffness matrix for three different resolutions. 

{
\begin{table}[ ] 
\caption{Material properties of a cracked Carrara marble sample}
\centering
\vspace{+1.0 mm}
 \begin{tabular}{| l | l | l |  }
  \hline			
  Material property & Background  & Pore space    \\
	\hline
	\hline
    Bulk modulus $K$       & 70\,GPa  & $10^{-3}$\,GPa   \\
  Shear modulus $G $   & 30\,GPa   & $2 \times 10^{-3}$\,GPa       \\
  \hline  
 \end{tabular}
 \label{tbl1real}
\end{table}
}

\section{Discussion}

\subsection{Accuracy of the numerical solver}

In the present study, the solution of the quasi-static (inertialess) equations is achieved via the pseudo-transient (PT) method, which is a version of a matrix-free iteration method. The accuracy of the PT method is of 1st order in time. Spatial discretization is performed using a 2nd order numerical scheme; however, since the time discretization is of the 1st order, the overall solution of the elasticity equations exhibits the 1st order accuracy.

Numerical tests show that even a single voxel element is sufficient to represent the crack thickness. Therefore, the present method can be used to calculate the effective elastic properties of materials represented by a low aspect ratio (e.g., $\alpha = 0.001$) spheroid (i.e., a crack) embedded into an isotropic matrix.

Another source of errors is related to the boundary conditions applied. In the present study, the velocity boundary conditions ($v = \Delta v$ on $\partial V$) are applied which is similar to applying a displacement boundary conditions ($u = \Delta u$ on $\partial V$). Therefore, the effective elastic properties are slightly stiffer compared to those obtained with periodic boundary conditions. Numerical tests show that if an inclusion is located in the center of the model and the distance to the model boundaries is about a diameter of the inclusion --- the effective properties of such a model will be close to the exact Eshelby solution with the cumulative error $< 0.5\%$. 

Another source of errors is related to the accuracy of the segmentation processing phase \citep{andra2013digital1, saxena2017effect}. Other sources of errors were discussed in \cite{garboczi1995algorithm}. A detailed discussion about other sources of errors presented in the numerical calculations  is outside of the scope of the present study.

\subsection{Performance}

The performance of the present solver depends on the GPU bandwidth and the numerical resolution. Overall, it can be concluded that the execution time is extremely fast, however, it depends on the GPU involved.

If a professional GPU (such as Nvidia A100 or newer) is used, the execution time can be 2-4 seconds for low-resolution models (up to 8 million voxel elements) or 2-3 minutes for high-resolution models (about 100 million voxel elements). The highest resolution that can fit into the 80GB DRAM GPU memory of A100 GPU is $703^3 \approx 350,000,000$ voxel elements and the computation of a single component of the stifness matrix takes 14 minutes. For a novel H200 GPU, the execution time is expected to be faster by $\times 3$. 

If a laptop GPU (such as Nvidia RTX 4090 or newer) is used, the execution time is approximately three times slower than that using an A100 GPU. For low-resolution models, it is about 10-12 seconds, and 3-4 minutes for moderate-resolution models (about 60 million voxel elements). In terms of the resolution, the only limitation is the available GPU DRAM memory.


\subsection{Future perspectives}

The presented application can be further extended in various ways. A few possible improvements are outlined below: \\

1. The present solver is developed to run on a single GPU. The solver can be extended to run on several GPUs in parallel via the Message Passing Interface (MPI) technology. It can be done for the present application following the methodology presented by \cite{alkhimenkov2021resolving}. Such an extension will allow us to run models with more than a billion voxel elements. Therefore, it will be possible to calculate effective properties for entire laboratory samples. \\

2. The present solver can be used to calculate anisotropic elastic properties up to orthorhombic symmetry (9 independent constants). The solver can be extended to calculate all 21 independent constants of the effective stiffness tensor.\\

3. In the present setup, the background material is isotropic. The application can be extended to deal with an anisotropic background. In the case of orthorhombic symmetry symmetry of the background, the extension is straightforward. If the target anisotropy of the background material is more anisotropic (i.e., 21 independent constants of $C_{mn}$), then the extension can be done but it will require averaging of some local fields or the modification of the numerical scheme.\\

4. In the present setup, the discretization is uniform. However, for some purposes, it might be useful to coarsen the discretization near the boundaries of the numerical domain and increase the discretization of the inclusion.\\


5. Due to fast execution times (2-4 seconds for 4-8 million voxel elements), the present solver can function as a main building block to resolve inverse problems.\\


6. The solver is designed to run on GPU devices. However, it is possible to simplify the solver to run purely in Matlab in 2D. The obvious disadvantage is that the resulting code will be much slower but it still might be useful to deal with small problems and for educational purposes and benchmarks.\\


\section{Conclusions}

An efficient and fast application for the calculation of effective elastic properties is provided. The application can be used to calculate effective properties of heterogeneous materials up to orthorhombic symmetry while it can be extended to deal with any symmetry class. The execution time of the program is less than 4 minutes on modern GPUs. The solver has been validated against exact analytical solutions. The solver has been applied to calculate the effective elastic properties of a 3-D digital image of a cracked Carrara marble sample. The application can be used to calculate effective elastic properties of heterogeneous materials at any scale up to 350 million voxel elements on a single professional GPU featuring 80GB memory. In the present solver, the maximum number of voxel elements is only restricted by the available GPU memory and can exceed 1 billion voxel elements.

\section{Acknowledgments}

The author is gratefully acknowledges support from the Swiss National Science Foundation, project number P500PN$\_$206722.

\section*{Computer Code Availability}
Routines that can be used to reproduce the presented results are available from \url{https://doi.org/10.5281/zenodo.13998253} \citep{alkhimenkov_2024_13998253} [Software].

\newpage

\textbf{Code availability section}

FastCijklElastic\_v1.1

Contact: yalkhime@mit.edu

Hardware requirements: Nvidia GPU 6GB DRAM

Program language: Matlab (GNU Octave), CUDA C
 
Software required: Matlab (GNU Octave), CUDA 12.5, C compiler


The source codes are available for downloading at the link:
\url{https://doi.org/10.5281/zenodo.13998253}


\appendix

\section{Boundary conditions}
\label{sec:sample:appendix}
 
For simplicity let us assume that the numerical domain $V$ is a cuboid $V = (0,Lx)\times(0,Ly)\times(0,Lz)$ and $\partial V$ its boundary $\partial V = \partial V^{xz0} \cup \partial V^{xzL} \cup \partial V^{yz0} \cup \partial V^{yzL} \cup \partial V^{xy0} \cup \partial V^{xyL}$, e.g., the boundary $\partial V^{xz0}$ represents a $xz$ plane with zero coordinate and $\partial V^{xzL}$ represents a $xz$ plane with $Ly$ coordinate. There are six planes in total. Note that here instead of $i,j,k =\overline{1..3}$, we employ $v_i$, where $i=x,y,z$.

\subsection{Normal compression test}

Normal compression direct test is adopted to calculate the $C_{33}$ component of the stiffness tensor:
 
 $\partial V^{xyL}$ is set to  $v_{z} =\Delta v$;  $v_{x},\,v_{y}$ are free (Dirichlet conditions)\\
 $\partial V^{xy0}$ is set to $v_{z} =0$; $v_{x} ,\, v_{y}$ are free (Dirichlet conditions)\\
 $\partial V^{xz0}$ and $\partial V^{xzL}$: $v_{x} ,\, v_{y} ,\, v_{z}$ are free (Dirichlet conditions)\\
 $\partial V^{yz0}$ and $\partial V^{yzL}$: $v_{x} ,\,v_{y},\, v_{z}$ are free (Dirichlet conditions),\\
 where $\Delta v = 10^{4}$.  Similar relaxation tests are performed to calculate $C_{11}$ and $C_{22}$ components of the stiffness matrix. The statement "are free" denotes that no special treatment of boundary conditions for the corresponding components is performed and the boundary conditions are implicitly imposed following the numerical discretization scheme (it is a standard reflecting boundary conditions for wave propagation).
 

\subsection{Simple shear test}

Simple shear relaxation tests are applied to calculate the components $C_{44}(\omega)$, $C_{55}(\omega)$ and $C_{66}(\omega)$. For example,  the $C_{55}$ ($xz$) component can be calculated using the following boundary conditions:

 $\partial V^{xyL}$ is set to $v_{x} =\Delta v$; $v_{z},v_{y}$ are free (Dirichlet conditions)\\
 $\partial V^{xy0}$ is set to $v_{x} =0$; $v_{z} =0,v_{z} =0$ are free (Dirichlet conditions)\\
 $\partial V^{xz0}$, $\partial V^{xzL}$, $\partial V^{yz0}$ and $\partial V^{yzL}$  are free-slip for all components.

\subsection{Direct tests for off-diagonal ($C_{13}$, $C_{12}$ and $C_{23}$) components}

The $C_{13}$, $C_{12}$ and $C_{23}$ components can be calculated by using the anisotropic stress-strain relation (Hooke's law) and appropriate boundary conditions:\\
$\partial V^{xyL}$ is set to  $v_{z} =\Delta v$;  $v_{x},v_{y}$ are free (Dirichlet conditions)\\
 $\partial V^{xy0}$ is set to $v_{z} =0$; $v_{x} ,v_{y}$ are free (Dirichlet conditions)\\
 
 $\partial V^{xzL}$ is set to  $v_{y} =\Delta v$;  $v_{x},v_{z}$ are free (Dirichlet conditions)\\
 $\partial V^{xz0}$ is set to $v_{y} =0$; $v_{x} ,v_{z}$ are free (Dirichlet conditions)\\
 
 $\partial V^{yzL}$ is set to  $v_{x} =\Delta v$;  $v_{y},v_{z}$ are free (Dirichlet conditions)\\
 $\partial V^{yz0}$ is set to $v_{x} =0$; $v_{y} ,v_{z}$ are free (Dirichlet conditions)\\

Let us write some useful expressions:
           \begin{equation}\label{D1}
     \bar{\sigma}_{ij} = \textbf{I}_2 \circ \sigma_k - (\textbf{I}_2 \circ C_k) \circ \varepsilon_k
    \end{equation} 
    and 
               \begin{equation}\label{D2}
     \bar{\varepsilon}_{ij} = \textbf{I}_2 \circ \varepsilon_k ,
    \end{equation} 
where $\sigma_k = (\langle \sigma_{11} \rangle, \langle\sigma_{22} \rangle, \langle\sigma_{33}) \rangle$, $C_k= (C_{11}, C_{22}, C_{33})$, $\varepsilon_k = (\langle \varepsilon_{11} \rangle,\langle \varepsilon_{22} \rangle, \langle \varepsilon_{33} \rangle)$, $\textbf{I}_2$ is the second-order identity tensor and $\circ$ denotes the Hadamard product (element-wise multiplication). Components $C_{33}$, $C_{22}$ and $C_{11}$ are taken from the (direct) normal compression tests. Let us write a system of equations for the off-diagonal components of the stiffness tensor (in Voigt notation) using expressions \eqref{D1}–\eqref{D2}, considering that $v_{x} = \Delta v$, $v_{y} = \Delta v$, and $v_{z} = \Delta v$ (which gives us three non-trivial components, allowing for the solution of three equations):
\begin{equation}
\left\{
\begin{aligned}
\bar{\varepsilon}_{22} C_{12} + \bar{\varepsilon}_{33} C_{13} &= \bar{\sigma}_{11}, \\
\bar{\varepsilon}_{11} C_{12} + \bar{\varepsilon}_{33} C_{23} &= \bar{\sigma}_{22}, \\
\bar{\varepsilon}_{11} C_{13} + \bar{\varepsilon}_{22} C_{23} &= \bar{\sigma}_{33}.
\end{aligned}
\right.
\label{D3v1}
\end{equation}
The system of equations \ref{D3v1} is solved analytically (symbolically) for the  off-diagonal components of the stiffness tensor. The system of equations \ref{D3v1} represents the stress-strain relation for all off-diagonal components of the stiffness tensor assuming non zero strains and stresses according to the boundary conditions described above. In 3D, similar direct tests for off-diagonal components were developed by \cite{alkhimenkov2020azimuth} (see Appendix A therein). The resulting expressions are:
           \begin{equation}\label{D3}
    C_{12} = \frac{ \bar{\varepsilon}_{11}   \bar{\sigma}_{11}  + \bar{\varepsilon}_{22}   \bar{\sigma}_{22} - \bar{\varepsilon}_{33}   \bar{\sigma}_{33} }{ 2 \bar{\varepsilon}_{11} \bar{\varepsilon}_{22} },
    \end{equation} 
           \begin{equation}\label{D4}
    C_{13} = \frac{ \bar{\varepsilon}_{11}   \bar{\sigma}_{11}  - \bar{\varepsilon}_{22}   \bar{\sigma}_{22} + \bar{\varepsilon}_{33}   \bar{\sigma}_{33} }{ 2 \bar{\varepsilon}_{11} \bar{\varepsilon}_{33} },
    \end{equation}
           \begin{equation}\label{D5}
    C_{23} = \frac{ -\bar{\varepsilon}_{11}   \bar{\sigma}_{11}  + \bar{\varepsilon}_{22}   \bar{\sigma}_{22} + \bar{\varepsilon}_{33}   \bar{\sigma}_{33} }{ 2 \bar{\varepsilon}_{22} \bar{\varepsilon}_{33} }.
    \end{equation}
Note that equations \ref{D3}-\ref{D5} are derived from the stress-strain relationship by considering non-zero strains in the $x$-, $y$-, and $z$-directions, followed by solving a system of three equations symbolically (analytically). A similar approach for calculating the off-diagonal components of the stiffness tensor can be found in \cite{carcione2011anisotropic, alkhimenkov2020frequency, alkhimenkov2020azimuth}.

\section{Property contribution tensor $\mathbf{H}$}
\label{sec:sample:appendix2}

The explicit form of the $\mathbf{H}$ tensor for an ellipsoidal inhomogeneity can be expressed as \citep{kachanov2018micromechanics}:
\begin{equation}\label{eqN21}
  \mathbf{H} = \left[ \left( \mathbf{S} ^{\textrm{2}} - \mathbf{S} ^{\textrm{1}}\right)^{-1} + \mathbf{Q} \right]^{-1}
\end{equation}
and 
\begin{equation}\label{eqN22}
\mathbf{Q} = \mathbf{C}^{\textrm{1}}: \left(\mathbf{I}_4  -  \mathbf{s} \right),
\end{equation}
where $\mathbf{C} ^{\textrm{1}} = \left( \mathbf{S} ^{\textrm{1}}  \right)^{-1}  $ is the stiffness tensor of the background material, $\mathbf{C} ^{\textrm{2}} = \left( \mathbf{S} ^{\textrm{2}}  \right)^{-1}  $ is the stiffness tensor of the inclusion, $\mathbf{s}$ is the 4-th rank Eshelby tensor \citep{eshelby1957determination}, and $\mathbf{I}_4$ is the 4-th rank identity tensor defined as
\begin{equation}\label{eqN23}
\mathbf{I}_4 = \dfrac{1}{2} \left( \delta_{ik} \delta_{jl}  +\delta_{il} \delta_{jk}   \right)
\end{equation}
The connection of tensor $\mathbf{Q}$ with the stress concentration tensor $\mathbf{\Gamma}$ is straightforward:
\begin{equation}\label{eqN24}
\mathbf{\Gamma} = \left[\mathbf{I}_4 + \mathbf{Q}:\left( \mathbf{S} ^{\textrm{2}} - \mathbf{S} ^{\textrm{1}}\right)      \right] ^{-1}, 
\end{equation}
where the stress concentration tensor gives the average stress in the inclusion with respect to the applied stress boundary condition. An alternative definition of the $\mathbf{H}$ tensor now can be written as 
\begin{equation}\label{eqN21}
  \mathbf{H} =  \left( \mathbf{S} ^{\textrm{2}} - \mathbf{S} ^{\textrm{1}}\right) : \mathbf{\Gamma}.
\end{equation}

\bibliographystyle{cas-model2-names}
\bibliography{cas-refs} 

\begin{thebibliography}{57}
\expandafter\ifx\csname natexlab\endcsname\relax\def\natexlab#1{#1}\fi
\providecommand{\url}[1]{\texttt{#1}}
\providecommand{\href}[2]{#2}
\providecommand{\path}[1]{#1}
\providecommand{\DOIprefix}{doi:}
\providecommand{\ArXivprefix}{arXiv:}
\providecommand{\URLprefix}{URL: }
\providecommand{\Pubmedprefix}{pmid:}
\providecommand{\doi}[1]{\href{http://dx.doi.org/#1}{\path{#1}}}
\providecommand{\Pubmed}[1]{\href{pmid:#1}{\path{#1}}}
\providecommand{\bibinfo}[2]{#2}
\ifx\xfnm\relax \def\xfnm[#1]{\unskip,\space#1}\fi
\bibitem[{Adam(2023)}]{AdamA}
\bibinfo{author}{Adam, A.}, \bibinfo{year}{2023}.
\newblock \bibinfo{title}{Mesh voxelisation}.
\newblock \URLprefix
  \url{https://www.mathworks.com/matlabcentral/fileexchange/27390-mesh-voxelisation}.
\bibitem[{Alkhimenkov(2024)}]{alkhimenkov_2024_13998253}
\bibinfo{author}{Alkhimenkov, Y.}, \bibinfo{year}{2024}.
\newblock \bibinfo{title}{Fast{C}ijkl{E}lastic}.
\newblock \URLprefix \url{https://doi.org/10.5281/zenodo.13998253},
  \DOIprefix\doi{10.5281/zenodo.13998253}.
\bibitem[{Alkhimenkov et~al.(2020a)Alkhimenkov, Caspari, Gurevich, Barbosa,
  Glubokovskikh, Hunziker and Quintal}]{alkhimenkov2020frequency}
\bibinfo{author}{Alkhimenkov, Y.}, \bibinfo{author}{Caspari, E.},
  \bibinfo{author}{Gurevich, B.}, \bibinfo{author}{Barbosa, N.D.},
  \bibinfo{author}{Glubokovskikh, S.}, \bibinfo{author}{Hunziker, J.},
  \bibinfo{author}{Quintal, B.}, \bibinfo{year}{2020}a.
\newblock \bibinfo{title}{Frequency-dependent attenuation and dispersion caused
  by squirt flow: Three-dimensional numerical study}.
\newblock \bibinfo{journal}{Geophysics} \bibinfo{volume}{85},
  \bibinfo{pages}{MR129--MR145}.
\bibitem[{Alkhimenkov et~al.(2020b)Alkhimenkov, Caspari, Lissa and
  Quintal}]{alkhimenkov2020azimuth}
\bibinfo{author}{Alkhimenkov, Y.}, \bibinfo{author}{Caspari, E.},
  \bibinfo{author}{Lissa, S.}, \bibinfo{author}{Quintal, B.},
  \bibinfo{year}{2020}b.
\newblock \bibinfo{title}{Azimuth-, angle-and frequency-dependent seismic
  velocities of cracked rocks due to squirt flow}.
\newblock \bibinfo{journal}{Solid Earth} \bibinfo{volume}{11},
  \bibinfo{pages}{855--871}.
\bibitem[{Alkhimenkov et~al.(2021a)Alkhimenkov, Khakimova and
  Podladchikov}]{alkhimenkov2021stability}
\bibinfo{author}{Alkhimenkov, Y.}, \bibinfo{author}{Khakimova, L.},
  \bibinfo{author}{Podladchikov, Y.}, \bibinfo{year}{2021}a.
\newblock \bibinfo{title}{Stability of discrete schemes of biot’s poroelastic
  equations}.
\newblock \bibinfo{journal}{Geophysical Journal International}
  \bibinfo{volume}{225}, \bibinfo{pages}{354--377}.
\bibitem[{Alkhimenkov et~al.(2024a)Alkhimenkov, Khakimova and
  Podladchikov}]{alkhimenkov2024shear}
\bibinfo{author}{Alkhimenkov, Y.}, \bibinfo{author}{Khakimova, L.},
  \bibinfo{author}{Podladchikov, Y.}, \bibinfo{year}{2024}a.
\newblock \bibinfo{title}{Shear bands triggered by solitary porosity waves in
  deforming fluid-saturated porous media}.
\newblock \bibinfo{journal}{Geophysical Research Letters} \bibinfo{volume}{51},
  \bibinfo{pages}{e2024GL108789}.
\bibitem[{Alkhimenkov et~al.(2024b)Alkhimenkov, Khakimova, Utkin and
  Podladchikov}]{https://doi.org/10.1029/2023JB028566}
\bibinfo{author}{Alkhimenkov, Y.}, \bibinfo{author}{Khakimova, L.},
  \bibinfo{author}{Utkin, I.}, \bibinfo{author}{Podladchikov, Y.},
  \bibinfo{year}{2024}b.
\newblock \bibinfo{title}{Resolving strain localization in frictional and
  time-dependent plasticity: Two- and three-dimensional numerical modeling
  study using graphical processing units (gpus)}.
\newblock \bibinfo{journal}{Journal of Geophysical Research: Solid Earth}
  \bibinfo{volume}{129}, \bibinfo{pages}{e2023JB028566}.
\newblock \URLprefix
  \url{https://agupubs.onlinelibrary.wiley.com/doi/abs/10.1029/2023JB028566},
  \DOIprefix\doi{https://doi.org/10.1029/2023JB028566},
  \href{http://arxiv.org/abs/https://agupubs.onlinelibrary.wiley.com/doi/pdf/10.1029/2023JB028566}{\tt
  arXiv:https://agupubs.onlinelibrary.wiley.com/doi/pdf/10.1029/2023JB028566}.
  \bibinfo{note}{e2023JB028566 2023JB028566}.
\bibitem[{Alkhimenkov and Podladchikov(2024)}]{alkhimenkov2024PT}
\bibinfo{author}{Alkhimenkov, Y.}, \bibinfo{author}{Podladchikov, Y.Y.},
  \bibinfo{year}{2024}.
\newblock \bibinfo{title}{Accelerated pseudo-transient method for elastic,
  viscoelastic, and coupled hydro-mechanical problems with applications}.
\newblock \bibinfo{journal}{Geoscientific Model Development Discussions}
  \bibinfo{volume}{2024}, \bibinfo{pages}{1--35}.
\bibitem[{Alkhimenkov et~al.(2021b)Alkhimenkov, R{\"a}ss, Khakimova, Quintal
  and Podladchikov}]{alkhimenkov2021resolving}
\bibinfo{author}{Alkhimenkov, Y.}, \bibinfo{author}{R{\"a}ss, L.},
  \bibinfo{author}{Khakimova, L.}, \bibinfo{author}{Quintal, B.},
  \bibinfo{author}{Podladchikov, Y.}, \bibinfo{year}{2021}b.
\newblock \bibinfo{title}{Resolving wave propagation in anisotropic poroelastic
  media using graphical processing units {(GPUs)}}.
\newblock \bibinfo{journal}{Journal of Geophysical Research: Solid Earth}
  \bibinfo{volume}{126}, \bibinfo{pages}{e2020JB021175}.
\bibitem[{Andr{\"a} et~al.(2013a)Andr{\"a}, Combaret, Dvorkin, Glatt, Han,
  Kabel, Keehm, Krzikalla, Lee, Madonna et~al.}]{andra2013digital1}
\bibinfo{author}{Andr{\"a}, H.}, \bibinfo{author}{Combaret, N.},
  \bibinfo{author}{Dvorkin, J.}, \bibinfo{author}{Glatt, E.},
  \bibinfo{author}{Han, J.}, \bibinfo{author}{Kabel, M.},
  \bibinfo{author}{Keehm, Y.}, \bibinfo{author}{Krzikalla, F.},
  \bibinfo{author}{Lee, M.}, \bibinfo{author}{Madonna, C.}, et~al.,
  \bibinfo{year}{2013}a.
\newblock \bibinfo{title}{Digital rock physics benchmarks—part i: Imaging and
  segmentation}.
\newblock \bibinfo{journal}{Computers \& Geosciences} \bibinfo{volume}{50},
  \bibinfo{pages}{25--32}.
\bibitem[{Andr{\"a} et~al.(2013b)Andr{\"a}, Combaret, Dvorkin, Glatt, Han,
  Kabel, Keehm, Krzikalla, Lee, Madonna et~al.}]{andra2013digital}
\bibinfo{author}{Andr{\"a}, H.}, \bibinfo{author}{Combaret, N.},
  \bibinfo{author}{Dvorkin, J.}, \bibinfo{author}{Glatt, E.},
  \bibinfo{author}{Han, J.}, \bibinfo{author}{Kabel, M.},
  \bibinfo{author}{Keehm, Y.}, \bibinfo{author}{Krzikalla, F.},
  \bibinfo{author}{Lee, M.}, \bibinfo{author}{Madonna, C.}, et~al.,
  \bibinfo{year}{2013}b.
\newblock \bibinfo{title}{Digital rock physics benchmarks—part ii: Computing
  effective properties}.
\newblock \bibinfo{journal}{Computers \& Geosciences} \bibinfo{volume}{50},
  \bibinfo{pages}{33--43}.
\bibitem[{Arns et~al.(2002)Arns, Knackstedt, Pinczewski and
  Garboczi}]{arns2002computation}
\bibinfo{author}{Arns, C.H.}, \bibinfo{author}{Knackstedt, M.A.},
  \bibinfo{author}{Pinczewski, W.V.}, \bibinfo{author}{Garboczi, E.J.},
  \bibinfo{year}{2002}.
\newblock \bibinfo{title}{Computation of linear elastic properties from
  microtomographic images: Methodology and agreement between theory and
  experiment}.
\newblock \bibinfo{journal}{Geophysics} \bibinfo{volume}{67},
  \bibinfo{pages}{1396--1405}.
\bibitem[{Backus(1962)}]{backus1962long}
\bibinfo{author}{Backus, G.E.}, \bibinfo{year}{1962}.
\newblock \bibinfo{title}{Long-wave elastic anisotropy produced by horizontal
  layering}.
\newblock \bibinfo{journal}{Journal of Geophysical Research}
  \bibinfo{volume}{67}, \bibinfo{pages}{4427--4440}.
\bibitem[{Bakhvalov and Panasenko(2012)}]{bakhvalov2012homogenisation}
\bibinfo{author}{Bakhvalov, N.S.}, \bibinfo{author}{Panasenko, G.},
  \bibinfo{year}{2012}.
\newblock \bibinfo{title}{Homogenisation: averaging processes in periodic
  media: mathematical problems in the mechanics of composite materials}.
  volume~\bibinfo{volume}{36}.
\newblock \bibinfo{publisher}{Springer Science \& Business Media}.
\bibitem[{Benveniste(1987)}]{benveniste1987new}
\bibinfo{author}{Benveniste, Y.}, \bibinfo{year}{1987}.
\newblock \bibinfo{title}{A new approach to the application of mori-tanaka's
  theory in composite materials}.
\newblock \bibinfo{journal}{Mechanics of materials} \bibinfo{volume}{6},
  \bibinfo{pages}{147--157}.
\bibitem[{Buryachenko(2001)}]{buryachenko2001multiparticle}
\bibinfo{author}{Buryachenko, V.}, \bibinfo{year}{2001}.
\newblock \bibinfo{title}{Multiparticle effective field and related methods in
  micromechanics of composite materials} .
\bibitem[{Carcione et~al.(2011)Carcione, Santos and
  Picotti}]{carcione2011anisotropic}
\bibinfo{author}{Carcione, J.}, \bibinfo{author}{Santos, J.E.},
  \bibinfo{author}{Picotti, S.}, \bibinfo{year}{2011}.
\newblock \bibinfo{title}{Anisotropic poroelasticity and wave-induced fluid
  flow: harmonic finite-element simulations}.
\newblock \bibinfo{journal}{Geophysical Journal International}
  \bibinfo{volume}{186}, \bibinfo{pages}{1245--1254}.
\bibitem[{Dormy and Tarantola(1995)}]{dormy1995numerical}
\bibinfo{author}{Dormy, E.}, \bibinfo{author}{Tarantola, A.},
  \bibinfo{year}{1995}.
\newblock \bibinfo{title}{Numerical simulation of elastic wave propagation
  using a finite volume method}.
\newblock \bibinfo{journal}{Journal of Geophysical Research: Solid Earth}
  \bibinfo{volume}{100}, \bibinfo{pages}{2123--2133}.
\bibitem[{Drach et~al.(2016)Drach, Tsukrov and Trofimov}]{drach2016comparison}
\bibinfo{author}{Drach, B.}, \bibinfo{author}{Tsukrov, I.},
  \bibinfo{author}{Trofimov, A.}, \bibinfo{year}{2016}.
\newblock \bibinfo{title}{Comparison of full field and single pore approaches
  to homogenization of linearly elastic materials with pores of regular and
  irregular shapes}.
\newblock \bibinfo{journal}{International Journal of Solids and Structures}
  \bibinfo{volume}{96}, \bibinfo{pages}{48--63}.
\bibitem[{Duretz et~al.(2019)Duretz, R{\"a}ss, Podladchikov and
  Schmalholz}]{duretz2019resolving}
\bibinfo{author}{Duretz, T.}, \bibinfo{author}{R{\"a}ss, L.},
  \bibinfo{author}{Podladchikov, Y.}, \bibinfo{author}{Schmalholz, S.},
  \bibinfo{year}{2019}.
\newblock \bibinfo{title}{Resolving thermomechanical coupling in two and three
  dimensions: spontaneous strain localization owing to shear heating}.
\newblock \bibinfo{journal}{Geophysical Journal International}
  \bibinfo{volume}{216}, \bibinfo{pages}{365--379}.
\bibitem[{Eshelby(1957)}]{eshelby1957determination}
\bibinfo{author}{Eshelby, J.D.}, \bibinfo{year}{1957}.
\newblock \bibinfo{title}{The determination of the elastic field of an
  ellipsoidal inclusion, and related problems}.
\newblock \bibinfo{journal}{Proceedings of the royal society of London. Series
  A. Mathematical and physical sciences} \bibinfo{volume}{241},
  \bibinfo{pages}{376--396}.
\bibitem[{Evstigneev et~al.(2023)Evstigneev, Ryabkov and
  Gerke}]{evstigneev2023stationary}
\bibinfo{author}{Evstigneev, N.M.}, \bibinfo{author}{Ryabkov, O.I.},
  \bibinfo{author}{Gerke, K.M.}, \bibinfo{year}{2023}.
\newblock \bibinfo{title}{Stationary stokes solver for single-phase flow in
  porous media: A blastingly fast solution based on algebraic multigrid method
  using gpu}.
\newblock \bibinfo{journal}{Advances in Water Resources} \bibinfo{volume}{171},
  \bibinfo{pages}{104340}.
\bibitem[{Frankel(1950)}]{frankel1950convergence}
\bibinfo{author}{Frankel, S.P.}, \bibinfo{year}{1950}.
\newblock \bibinfo{title}{Convergence rates of iterative treatments of partial
  differential equations}.
\newblock \bibinfo{journal}{Mathematics of Computation} \bibinfo{volume}{4},
  \bibinfo{pages}{65--75}.
\bibitem[{Garboczi(1998)}]{garboczi1998finite}
\bibinfo{author}{Garboczi, E.J.}, \bibinfo{year}{1998}.
\newblock \bibinfo{title}{Finite element and finite difference programs for
  computing the linear electric and elastic properties of digital images of
  random materials} .
\bibitem[{Garboczi and Day(1995)}]{garboczi1995algorithm}
\bibinfo{author}{Garboczi, E.J.}, \bibinfo{author}{Day, A.R.},
  \bibinfo{year}{1995}.
\newblock \bibinfo{title}{An algorithm for computing the effective linear
  elastic properties of heterogeneous materials: three-dimensional results for
  composites with equal phase poisson ratios}.
\newblock \bibinfo{journal}{Journal of the Mechanics and Physics of Solids}
  \bibinfo{volume}{43}, \bibinfo{pages}{1349--1362}.
\bibitem[{Kachanov and Sevostianov(2018)}]{kachanov2018micromechanics}
\bibinfo{author}{Kachanov, M.}, \bibinfo{author}{Sevostianov, I.},
  \bibinfo{year}{2018}.
\newblock \bibinfo{title}{Micromechanics of materials, with applications}.
  volume \bibinfo{volume}{249}.
\newblock \bibinfo{publisher}{Springer}.
\bibitem[{Landau and Lifshitz(1959)}]{landau1959courseElast}
\bibinfo{author}{Landau, L.D.}, \bibinfo{author}{Lifshitz, E.M.},
  \bibinfo{year}{1959}.
\newblock \bibinfo{title}{Course of Theoretical Physics Vol 7: Theory of
  Elasticity}.
\newblock \bibinfo{publisher}{Pergamon press}.
\bibitem[{Levin et~al.(2019)Levin, Zingerman, Yakovlev, Kurdenkova and
  Nemtinova}]{levin2019numerical}
\bibinfo{author}{Levin, V.A.}, \bibinfo{author}{Zingerman, K.M.},
  \bibinfo{author}{Yakovlev, M.Y.}, \bibinfo{author}{Kurdenkova, E.O.},
  \bibinfo{author}{Nemtinova, D.V.}, \bibinfo{year}{2019}.
\newblock \bibinfo{title}{Numerical estimation of effective properties of
  periodic cellular structures using beam and shell finite elements with cae
  fidesys}.
\newblock \bibinfo{journal}{Chebyshevskii Sbornik} \bibinfo{volume}{20},
  \bibinfo{pages}{523--536}.
\bibitem[{Li and Wang(2008)}]{li2008introduction}
\bibinfo{author}{Li, S.}, \bibinfo{author}{Wang, G.}, \bibinfo{year}{2008}.
\newblock \bibinfo{title}{Introduction to micromechanics and nanomechanics}.
\newblock \bibinfo{publisher}{World Scientific}.
\bibitem[{Lissa et~al.(2021a)Lissa, Ruf, Steeb and Quintal}]{lissa2021digital}
\bibinfo{author}{Lissa, S.}, \bibinfo{author}{Ruf, M.}, \bibinfo{author}{Steeb,
  H.}, \bibinfo{author}{Quintal, B.}, \bibinfo{year}{2021}a.
\newblock \bibinfo{title}{Digital rock physics applied to squirt flow}.
\newblock \bibinfo{journal}{Geophysics} \bibinfo{volume}{86},
  \bibinfo{pages}{MR235--MR245}.
\bibitem[{Lissa et~al.(2021b)Lissa, Ruf, Steeb and
  Quintal}]{simon_lissa_2021_4746139}
\bibinfo{author}{Lissa, S.}, \bibinfo{author}{Ruf, M.}, \bibinfo{author}{Steeb,
  H.}, \bibinfo{author}{Quintal, B.}, \bibinfo{year}{2021}b.
\newblock \bibinfo{title}{Digital rock physics applied to squirt flow}.
\newblock \URLprefix \url{https://doi.org/10.5281/zenodo.4746139},
  \DOIprefix\doi{10.5281/zenodo.4746139}.
\bibitem[{Markov et~al.(2020)Markov, Trofimov and
  Sevostianov}]{markov2020unified}
\bibinfo{author}{Markov, A.}, \bibinfo{author}{Trofimov, A.},
  \bibinfo{author}{Sevostianov, I.}, \bibinfo{year}{2020}.
\newblock \bibinfo{title}{A unified methodology for calculation of compliance
  and stiffness contribution tensors of inhomogeneities of arbitrary 2d and 3d
  shapes embedded in isotropic matrix--open access software.}
\newblock \bibinfo{journal}{International Journal of Engineering Science}
  \bibinfo{volume}{157}, \bibinfo{pages}{103390}.
\bibitem[{MeshLib(2023)}]{MeshLib}
\bibinfo{author}{MeshLib}, \bibinfo{year}{2023}.
\newblock \bibinfo{title}{https://github.com/meshinspector/meshlib}.
\newblock \URLprefix \url{https://github.com/MeshInspector/MeshLib}.
\bibitem[{Mori and Tanaka(1973)}]{mori1973average}
\bibinfo{author}{Mori, T.}, \bibinfo{author}{Tanaka, K.}, \bibinfo{year}{1973}.
\newblock \bibinfo{title}{Average stress in matrix and average elastic energy
  of materials with misfitting inclusions}.
\newblock \bibinfo{journal}{Acta metallurgica} \bibinfo{volume}{21},
  \bibinfo{pages}{571--574}.
\bibitem[{Moulinec and Suquet(1998)}]{moulinec1998numerical}
\bibinfo{author}{Moulinec, H.}, \bibinfo{author}{Suquet, P.},
  \bibinfo{year}{1998}.
\newblock \bibinfo{title}{A numerical method for computing the overall response
  of nonlinear composites with complex microstructure}.
\newblock \bibinfo{journal}{Computer methods in applied mechanics and
  engineering} \bibinfo{volume}{157}, \bibinfo{pages}{69--94}.
\bibitem[{Multiphysics(1998)}]{multiphysics1998introduction}
\bibinfo{author}{Multiphysics, C.}, \bibinfo{year}{1998}.
\newblock \bibinfo{title}{Introduction to comsol
  multiphysics{\textregistered}}.
\newblock \bibinfo{journal}{COMSOL Multiphysics, Burlington, MA, accessed Feb}
  \bibinfo{volume}{9}, \bibinfo{pages}{2018}.
\bibitem[{Nemat-Nasser and Hori(2013)}]{nemat2013micromechanics}
\bibinfo{author}{Nemat-Nasser, S.}, \bibinfo{author}{Hori, M.},
  \bibinfo{year}{2013}.
\newblock \bibinfo{title}{Micromechanics: overall properties of heterogeneous
  materials}.
\newblock \bibinfo{publisher}{Elsevier}.
\bibitem[{Omlin et~al.(2017)Omlin, Malvoisin and Podladchikov}]{omlin2017pore}
\bibinfo{author}{Omlin, S.}, \bibinfo{author}{Malvoisin, B.},
  \bibinfo{author}{Podladchikov, Y.Y.}, \bibinfo{year}{2017}.
\newblock \bibinfo{title}{Pore fluid extraction by reactive solitary waves in
  3-d}.
\newblock \bibinfo{journal}{Geophysical Research Letters} \bibinfo{volume}{44},
  \bibinfo{pages}{9267--9275}.
\bibitem[{Omlin et~al.(2018)Omlin, R{\"a}ss and
  Podladchikov}]{omlin2018simulation}
\bibinfo{author}{Omlin, S.}, \bibinfo{author}{R{\"a}ss, L.},
  \bibinfo{author}{Podladchikov, Y.Y.}, \bibinfo{year}{2018}.
\newblock \bibinfo{title}{Simulation of three-dimensional viscoelastic
  deformation coupled to porous fluid flow}.
\newblock \bibinfo{journal}{Tectonophysics} \bibinfo{volume}{746},
  \bibinfo{pages}{695--701}.
\bibitem[{Otero et~al.(2018)Otero, Oller and Martinez}]{otero2018multiscale}
\bibinfo{author}{Otero, F.}, \bibinfo{author}{Oller, S.},
  \bibinfo{author}{Martinez, X.}, \bibinfo{year}{2018}.
\newblock \bibinfo{title}{Multiscale computational homogenization: review and
  proposal of a new enhanced-first-order method}.
\newblock \bibinfo{journal}{Archives of Computational Methods in Engineering}
  \bibinfo{volume}{25}, \bibinfo{pages}{479--505}.
\bibitem[{Papanicolau et~al.(1978)Papanicolau, Bensoussan and
  Lions}]{papanicolau1978asymptotic}
\bibinfo{author}{Papanicolau, G.}, \bibinfo{author}{Bensoussan, A.},
  \bibinfo{author}{Lions, J.L.}, \bibinfo{year}{1978}.
\newblock \bibinfo{title}{Asymptotic analysis for periodic structures}.
\newblock \bibinfo{publisher}{Elsevier}.
\bibitem[{R{\"a}ss et~al.(2019)R{\"a}ss, Duretz and
  Podladchikov}]{rass2019resolving}
\bibinfo{author}{R{\"a}ss, L.}, \bibinfo{author}{Duretz, T.},
  \bibinfo{author}{Podladchikov, Y.}, \bibinfo{year}{2019}.
\newblock \bibinfo{title}{Resolving hydromechanical coupling in two and three
  dimensions: spontaneous channelling of porous fluids owing to decompaction
  weakening}.
\newblock \bibinfo{journal}{Geophysical Journal International}
  \bibinfo{volume}{218}, \bibinfo{pages}{1591--1616}.
\bibitem[{R{\"a}ss et~al.(2020)R{\"a}ss, Licul, Herman, Podladchikov and
  Suckale}]{rass2020modelling}
\bibinfo{author}{R{\"a}ss, L.}, \bibinfo{author}{Licul, A.},
  \bibinfo{author}{Herman, F.}, \bibinfo{author}{Podladchikov, Y.Y.},
  \bibinfo{author}{Suckale, J.}, \bibinfo{year}{2020}.
\newblock \bibinfo{title}{Modelling thermomechanical ice deformation using an
  implicit pseudo-transient method (fastice v1. 0) based on graphical
  processing units (gpus)}.
\newblock \bibinfo{journal}{Geoscientific Model Development}
  \bibinfo{volume}{13}, \bibinfo{pages}{955--976}.
\bibitem[{R{\"a}ss et~al.(2022)R{\"a}ss, Utkin, Duretz, Omlin and
  Podladchikov}]{rass2022assessing}
\bibinfo{author}{R{\"a}ss, L.}, \bibinfo{author}{Utkin, I.},
  \bibinfo{author}{Duretz, T.}, \bibinfo{author}{Omlin, S.},
  \bibinfo{author}{Podladchikov, Y.Y.}, \bibinfo{year}{2022}.
\newblock \bibinfo{title}{Assessing the robustness and scalability of the
  accelerated pseudo-transient method}.
\newblock \bibinfo{journal}{Geoscientific Model Development}
  \bibinfo{volume}{15}, \bibinfo{pages}{5757--5786}.
\bibitem[{Roberts and Garboczi(2002)}]{roberts2002computation}
\bibinfo{author}{Roberts, A.P.}, \bibinfo{author}{Garboczi, E.J.},
  \bibinfo{year}{2002}.
\newblock \bibinfo{title}{Computation of the linear elastic properties of
  random porous materials with a wide variety of microstructure}.
\newblock \bibinfo{journal}{Proceedings of the Royal Society of London. Series
  A: Mathematical, Physical and Engineering Sciences} \bibinfo{volume}{458},
  \bibinfo{pages}{1033--1054}.
\bibitem[{Ruf and Steeb(2020)}]{ruf2020micro}
\bibinfo{author}{Ruf, M.}, \bibinfo{author}{Steeb, H.}, \bibinfo{year}{2020}.
\newblock \bibinfo{title}{micro-xrct data set of carrara marble with
  artificially created crack network: fast cooling down from 600° c}.
\newblock \bibinfo{journal}{DaRUS, https://doi. org/10.18419/DARUS-682} .
\bibitem[{Saenger(2008)}]{saenger2008numerical}
\bibinfo{author}{Saenger, E.H.}, \bibinfo{year}{2008}.
\newblock \bibinfo{title}{Numerical methods to determine effective elastic
  properties}.
\newblock \bibinfo{journal}{International Journal of Engineering Science}
  \bibinfo{volume}{46}, \bibinfo{pages}{598--605}.
\bibitem[{Saenger et~al.(2007)Saenger, Ciz, Kr{\"u}ger, Schmalholz, Gurevich
  and Shapiro}]{saenger2007finite}
\bibinfo{author}{Saenger, E.H.}, \bibinfo{author}{Ciz, R.},
  \bibinfo{author}{Kr{\"u}ger, O.S.}, \bibinfo{author}{Schmalholz, S.M.},
  \bibinfo{author}{Gurevich, B.}, \bibinfo{author}{Shapiro, S.A.},
  \bibinfo{year}{2007}.
\newblock \bibinfo{title}{Finite-difference modeling of wave propagation on
  microscale: A snapshot of the work in progress}.
\newblock \bibinfo{journal}{Geophysics} \bibinfo{volume}{72},
  \bibinfo{pages}{SM293--SM300}.
\bibitem[{S{\'a}nchez-Palencia(1980)}]{sanchez1980non}
\bibinfo{author}{S{\'a}nchez-Palencia, E.}, \bibinfo{year}{1980}.
\newblock \bibinfo{title}{Non-homogeneous media and vibration theory}.
\newblock \bibinfo{journal}{Lecture Note in Physics, Springer-Verlag}
  \bibinfo{volume}{320}, \bibinfo{pages}{57--65}.
\bibitem[{Saxena et~al.(2017)Saxena, Hofmann, Alpak, Dietderich, Hunter and
  Day-Stirrat}]{saxena2017effect}
\bibinfo{author}{Saxena, N.}, \bibinfo{author}{Hofmann, R.},
  \bibinfo{author}{Alpak, F.O.}, \bibinfo{author}{Dietderich, J.},
  \bibinfo{author}{Hunter, S.}, \bibinfo{author}{Day-Stirrat, R.J.},
  \bibinfo{year}{2017}.
\newblock \bibinfo{title}{Effect of image segmentation \& voxel size on
  micro-ct computed effective transport \& elastic properties}.
\newblock \bibinfo{journal}{Marine and Petroleum Geology} \bibinfo{volume}{86},
  \bibinfo{pages}{972--990}.
\bibitem[{Saxena and Mavko(2016)}]{saxena2016estimating}
\bibinfo{author}{Saxena, N.}, \bibinfo{author}{Mavko, G.},
  \bibinfo{year}{2016}.
\newblock \bibinfo{title}{Estimating elastic moduli of rocks from thin
  sections: Digital rock study of 3d properties from 2d images}.
\newblock \bibinfo{journal}{Computers \& Geosciences} \bibinfo{volume}{88},
  \bibinfo{pages}{9--21}.
\bibitem[{Schneider(2021)}]{schneider2021review}
\bibinfo{author}{Schneider, M.}, \bibinfo{year}{2021}.
\newblock \bibinfo{title}{A review of nonlinear fft-based computational
  homogenization methods}.
\newblock \bibinfo{journal}{Acta Mechanica} \bibinfo{volume}{232},
  \bibinfo{pages}{2051--2100}.
\bibitem[{Shermergor(1977)}]{shermergor1977Theory}
\bibinfo{author}{Shermergor, T.D.}, \bibinfo{year}{1977}.
\newblock \bibinfo{title}{Theory of elasticity of micro-heterogeneous media (in
  Russian)}.
\newblock \bibinfo{publisher}{Nauka}.
\bibitem[{Virieux(1986)}]{virieux1986p}
\bibinfo{author}{Virieux, J.}, \bibinfo{year}{1986}.
\newblock \bibinfo{title}{P-sv wave propagation in heterogeneous media:
  Velocity-stress finite-difference method}.
\newblock \bibinfo{journal}{Geophysics} \bibinfo{volume}{51},
  \bibinfo{pages}{889--901}.
\bibitem[{Willis(1983)}]{willis1983overall}
\bibinfo{author}{Willis, J.R.}, \bibinfo{year}{1983}.
\newblock \bibinfo{title}{The overall elastic response of composite materials}
  .
\bibitem[{Zaoui(2002)}]{zaoui2002continuum}
\bibinfo{author}{Zaoui, A.}, \bibinfo{year}{2002}.
\newblock \bibinfo{title}{Continuum micromechanics: survey}.
\newblock \bibinfo{journal}{Journal of Engineering Mechanics}
  \bibinfo{volume}{128}, \bibinfo{pages}{808--816}.
\bibitem[{Zhang et~al.(2007)Zhang, Dai, Wang, Sun and Bassir}]{zhang2007using}
\bibinfo{author}{Zhang, W.}, \bibinfo{author}{Dai, G.}, \bibinfo{author}{Wang,
  F.}, \bibinfo{author}{Sun, S.}, \bibinfo{author}{Bassir, H.},
  \bibinfo{year}{2007}.
\newblock \bibinfo{title}{Using strain energy-based prediction of effective
  elastic properties in topology optimization of material microstructures}.
\newblock \bibinfo{journal}{Acta Mechanica Sinica} \bibinfo{volume}{23},
  \bibinfo{pages}{77--89}.

\end{thebibliography}

\end{document}